\documentclass[12pt]{article}
\usepackage{euscript,amsmath, amssymb, amsfonts}

\newcommand{\R}{{\mathbb R}}
\newcommand{\G}{{\mathbb G}}
\newcommand{\K}{{\mathbb K}}

\newcommand{\Z}{{\mathbb Z}}
\newcommand{\oC}{{\mathbb C}}

\newcommand{\di}{{\rm d}}

\newcommand{\p}{{\hbar^2}}
\newcommand{\pp}[2]{\hbar^{#1}}
\newcounter{sectionn}
\newcounter{theorem}
\newcounter{lemma}
\newcounter{proposition}
\newcounter{defin}
\newcounter{remark}

\newcommand{\sectionn}[1]{\par\noindent\refstepcounter{sectionn}
                             {\bf \arabic{sectionn}. #1.}}
\begin{document}
\sloppy
\title
 {
            \vspace{1cm}
       \textbf{General form of deformation of Poisson superbracket}
 }
\author
 {
 S.E.~Konstein\thanks{E-mail: konstein@lpi.ru}~,
~A.G.~Smirnov{\thanks{E-mail: smirnov@lpi.ru}}
~~and~I.V.~Tyutin\thanks{E-mail: tyutin@lpi.ru}
\\ {\small
               \phantom{uuu}} \\ {\small I.E.Tamm Department of
               Theoretical Physics,} \\
               {\small P. N. Lebedev Physical
               Institute,} \\ {\small 119991, Leninsky Prospect 53,
               Moscow, Russia.} }

\date{}
 \maketitle

\begin{abstract}
{ \footnotesize
Continuous formal deformations of the Poisson superbracket defined
on compactly supported smooth functions on $\R^n$
taking values in a Grassmann algebra are described up to an
 equivalence
transformation.
It is shown that 
there are the additional
deformations which are different from the standard Moyal
bracket.}
\end{abstract}

\sectionn{Introduction}
The hope to construct the quantum mechanics on nontrivial
manifolds is connected with geometrical or deformation
quantization \cite{1} - \cite{4}.
The functions on the phase space
are associated with the operators, and the product and the
commutator of the operators are described by associative *-product
and *-commutator of the functions. These *-product and
*-commutator are the deformations of usual product and of
usual Poisson bracket.
In \cite{Leites}, the problem is considered for the
Poisson superalgebra on the superspace of polynomials and for the antibracket.
Purely Grassmannian case is considered in \cite{Ty2} and
purely bosonic case was
considered in \cite{Batalin} under some additional assumptions.

In the present paper, we find the general form of the $*$-commutator in the
case of a Poisson superalgebra of smooth compactly supported functions taking
values in a Grassmann algebra.
The proposed analysis is essentially based on
the results of the paper~\cite{Cohom} by the authors, where the second
cohomology space with coefficients in the adjoint representation of the
Poisson superalgebra was found.

\sectionn{Deformations of topological Lie superalgebras}
In this section, we
recall some concepts concerning formal deformations of algebras (see,
e.g.,~\cite{Gerstenhaber}), adapting them to the case of topological Lie
superalgebras.
Let $L$ be a topological Lie superalgebra over $\K$ ($\K=\R$
or $\oC$) with Lie superbracket $\{\cdot,\cdot\}$, $\K[[\p]]$ be the ring of
formal power series in $\p$ over $\K$, and $L[[\p]]$ be the $\K[[\p]]$-module
of formal power series in $\p$ with coefficients in $L$.
We endow both
$\K[[\p]]$ and $L[[\p]]$ by the direct-product topology.
The grading of $L$
naturally determines a grading of $L[[\p]]$:  an element $f=f_0+\p
f_1+\ldots$ has a definite parity $\varepsilon(f)$ if
$\varepsilon(f)=\varepsilon(f_j)$ for all $j=0,1,..$.
Every $p$-linear
separately continuous mapping from $L^p$ to $L$ (in particular, the bracket
$\{\cdot,\cdot\}$) is uniquely extended by $\K[[\p]]$-linearity to a
$p$-linear separately continuous mapping over $\K[[\p]]$ from $L[[\p]]^p$ to
$L[[\p]]$.
A (continuous) formal deformation of $L$ is by definition a
$\K[[\p]]$-bilinear separately continuous Lie superbracket $C(\cdot,\cdot)$
on $L[[\p]]$ such that $C(f,g)=\{f,g\} \mod \p$ for any $f,g\in L[[\p]]$.
Obviously, every formal deformation $C$ is expressible in the form
\begin{equation}\label{1} C(f,g)=\{f,g\}+\p
C_1(f,g)+\pp{4}{2}C_2(f,g)+\ldots,\quad f,g\in L, \end{equation} where $C_j$
are separately continuous skew-symmetric bilinear mappings from $L\times L$
to $L$ (2-cochains with coefficients in the adjoint representation of $L$).
Formal deformations $C^1$ and $C^2$ are called equivalent if there is a
continuous $\K[[\p]]$-linear operator $T: L[[\p]]\to L[[\p]]$ such that
$TC^1(f,g)=C^2(T f,Tg)$, $f,g\in L[[\p]]$.  The problem of finding formal
deformations of $L$ is closely related to the problem of computing
Chevalle--Eilenberg cohomology of $L$ with coefficients in the adjoint
representation of $L$.  Let $\mathcal C_p(L)$ denote the space of $p$-linear
skew-symmetric separately continuous mappings from $L^p$ to $L$ (the space of
$p$-cochains with coefficients in the adjoint representation of $L$).  The
space $\mathcal C_p(L)$ possesses a natural $\Z_2$-grading: by definition,
$M_p\in \mathcal C_p(L)$ has the definite parity $\epsilon(M_p)$ if the
relation
$$
\varepsilon(M_p(f_1,\ldots,f_p))=
\varepsilon(M_p)+\varepsilon(f_1)+\ldots+\varepsilon(f_1)
$$
holds for any $f_j\in L$ with definite parities $\epsilon(f_j)$.
Since a Lie superbracket is always even,
all $C_j$ in the expansion~(\ref{1}) should be even 2-cochains.
The differential $\di_p^{\rm ad}$ is defined to be the linear
operator from $\mathcal C_p(L)$ to $\mathcal C_{p+1}(L)$ such that
\begin{align}
&&d_p^{\rm ad}M_p(f_1,...,f_{p+1})=
-\sum_{j=1}^{p+1}(-1)^{j+\varepsilon(f_j)|\varepsilon(f)|_{1,j-1}+
\varepsilon(f_j)\varepsilon_{M_p}}\{f_j,
M_p(f_{1},...,\hat{f}_j,...,f_{p+1})\}- \nonumber \\
&&-\sum_{i<j}(-1)^{j+\varepsilon(f_j)|\varepsilon(f)|_{i+1,j-1}}
M_p(f_1,...f_{i-1},\{f_i,f_j\},f_{i+1},...,\hat{f}_j,...,f_{p+1}),\label{diff}
\end{align}
for any $M_p\in \mathcal C_p(L)$ and $f_1,\ldots f_{p+1}\in L$
having definite parities.
Here the hat means that the argument is omitted and the notation
$$
|\varepsilon(f)|_{i,j}=\sum_{l=i}^j\varepsilon(f_l)
$$
has been used. Writing the Jacobi identity
for a deformation $C$ of the form~(\ref{1}),
\begin{equation}\label{6.1a}
\sigma(f,h)C(f,C(g,h))+\mathrm{cycle}(f,g,h)=0,
\quad \sigma(f,g)=(-1)^{\varepsilon(f)\varepsilon(g)},
\end{equation}
and taking the terms of the order $\p$, we find that
\begin{equation}\label{1a}
d_2^{\rm ad}C_1=0.
\end{equation}
Thus, the
first order deformations of $L$
are described by 2-cocycles of the differential $d^{\rm ad}$.

\sectionn{Poisson superalgebra}
Let $\EuScript D(\R^k)$ denote the space of
smooth $\K$-valued functions with compact support on $\R^k$.
This space is
endowed by its standard topology: by definition, a sequence $\varphi_k\in
\EuScript D(\R^k)$ converges to $\varphi\in \EuScript D(\R^k)$ if
$\partial^\lambda\varphi_k$ converge uniformly to $\partial^\lambda\varphi$
for every multi-index $\lambda$, and the supports of all $\varphi_k$ are
contained in a fixed compact set. We set
$$ \mathbf D^{n_-}_{n_+}= \EuScript
D(\R^{n_+})\otimes \G^{n_-},\quad \mathbf E^{n_-}_{n_+}=
C^\infty(\R^{n_+})\otimes \G^{n_-},
$$
where $\G^{n_-}$ is the Grassmann
algebra with $n_-$ generators and $\EuScript D'(\R^{n_+})$ is the space of
continuous linear functionals on $\EuScript D(\R^{n_+})$.  The generators of
the Grassmann algebra (resp., the coordinates of the space $\R^{n_+}$) are
denoted by $\xi^\alpha$, $\alpha=1,\ldots,n_-$ (resp., $x^i$, $i=1,\ldots,
n_+$).  We shall also use collective variables $z^A$ which are equal to $x^A$
for $A=1,\ldots,n_+$ and are equal to $\xi^{A-n_+}$ for
$A=n_++1,\ldots,n_++n_-$.  The spaces $\mathbf D^{n_-}_{n_+}$, $\mathbf
E^{n_-}_{n_+}$, and $\mathbf D^{\prime n_-}_{n_+}$ possess a natural grading
which is determined by that of the Grassmann algebra. The parity of an
element $f$ of these spaces is denoted by $\varepsilon(f)$. We also set
$\varepsilon_A=0$ for $A=1,\ldots, n_+$ and $\varepsilon_A=1$ for
$A=n_++1,\ldots, n_++n_-$.

Let $\partial/\partial z^A$ and $\overleftarrow{\partial}/\partial z^A$ be
the operators of the left and right differentiation.  The Poisson bracket is
defined by the relation \begin{equation}
\{f,g\}(z)=f(z)\frac{\overleftarrow{\partial}}{\partial z^A}\omega^{AB}
\frac{\partial}{\partial z^B}g(z)= -\sigma(f,g)\{g,f\}(z),\label{3.0}
\end{equation}
where
the symplectic metric $\omega^{AB}=(-1)^{\varepsilon_A
\varepsilon_B}\omega^{BA}$ is a constant invertible matrix.  For
 definiteness, we choose it in the form
\[
\omega^{AB}=
\left(\begin{array}{cc}\omega^{ij}&0       \\
0&\lambda_\alpha\delta^{\alpha\beta}\end{array}\right),\quad
\lambda_\alpha=\pm1,\ i,j=1,...,n_+,\ \alpha,\beta=1,...,n_-
\]
 where
$\omega^{ij}$ is the canonical symplectic form (if $\K=\oC$, then one can
choose $\lambda_\alpha=1$).  The nondegeneracy of the matrix $\omega^{AB}$
implies, in particular, that $n_+$ is even.  The Poisson superbracket
satisfies the Jacobi identity \begin{equation}
\sigma(f,h)\{f,\{g,h\}\}(z)+\hbox{cycle}(f,g,h)= 0,\quad f,g,h\in \mathbf
E^{n_-}_{n_+}.  \label{3.0a} \end{equation} By Poisson superalgebra $\mathcal
P$, we mean the space $\mathbf D^{n_-}_{n_+}$ with the Poisson
bracket~(\ref{3.0}) on it.  The relations~(\ref{3.0}) and~(\ref{3.0a}) show
that this bracket indeed determines a Lie superalgebra structure on $\mathbf
D^{n_-}_{n_+}$.

The integral on $\mathbf D^{n_-}_{n_+}$ is defined by the relation
$$
\bar
f\stackrel{\mathrm{def}}{=}\int \di z\, f(z)= \int_{\R^{n_+}}\di x\int
\di\xi\, f(z),
$$
where the integral on the Grassmann algebra is normed by
the condition $\int \di\xi\, \xi^1\ldots\xi^{n_-}=1$.

\sectionn{Formulation of the results} For any $\kappa\in \K[[\p]]$, the
Moyal-type superbracket
\begin{equation}\label{2} {\cal
M}_\kappa(z|f,g)=\frac{1}{\hbar\kappa}f(z)\sinh
\left(\hbar\kappa\frac{\overleftarrow{\partial}}{\partial z^A}\omega^{AB}
\frac{\partial}{\partial z^B}\right)g(z)
\end{equation}
is skew-symmetric and
satisfies the Jacobi identity
\footnote{ This fact is well known for the bosonic case
and can be easily checked for the supercase.}
and, therefore, gives a deformation of the
initial Poisson algebra. For $\zeta\in \mathbf E^{n_-}_{n_+}[[\p]]$, $\kappa,
c\in \K[[\p]]$, we set \begin{align} &{\cal N}_{\kappa,\zeta}(z|f,g)= {\cal
M}_\kappa(z|f+\zeta\bar{f},g+\zeta\bar{g}),\nonumber\\
&Q_\zeta(z|f,g)=\{f(z),\zeta(z)\}\bar{g}-
\sigma(f,g)\{g(z),\zeta(z)\}\bar{f},\label{Q}\\
&R(z|f,g)=(1-\frac{1}{2}z^A\frac{\partial}{\partial z^A})f(z)\bar{g}-
\sigma(f,g)(1-\frac{1}{2}z^A\frac{\partial}{\partial
z^A})g(z)\bar{f},\label{R}\\ &{\cal N}_{\kappa,\zeta,c}(z|f,g)= {\cal
M}_\kappa(z|f+\zeta\bar{f},g+\zeta\bar{g})+c\bar{f}\bar{g}\nonumber
\end{align}

Now we can formulate the main result of the present paper.

\medskip\noindent
{\textbf{Theorem.}
\begin{enumerate}
\item Let $n_-=2k$. Then every
continuous formal deformation of the Poisson superalgebra $\mathcal P$ is
equivalent either to the superbracket ${\cal N}_{\kappa,\zeta}(z|f,g)$, where
$\zeta\in \mathbf E^{n_-}_{n_+}[[\p]]$ is even and $\kappa\in \K[[\p]]$, or
to the superbracket
$$
C(z|f,g)=\{f(z),g(z)\}+Q_{\zeta}(z|f,g)+ c R(z|f,g),
$$
where $\zeta\in \mathbf E^{n_-}_{n_+}[[\p]]$ is even and $c\in \K[[\p]]$.

\item Let $n_-=2k+1$. Then every continuous formal deformation of the Poisson
superalgebra $\mathcal P$ is equivalent to the superbracket ${\cal
N}_{\kappa,\zeta,c}(z|f,g)$, where $c,\kappa\in \K[[\p]]$ and $\zeta\in
\mathbf E^{n_-}_{n_+}[[\p]]$ is an odd function such that $[{\cal
M}_{\kappa}(z|\zeta,\zeta)+ c]\in \mathbf D^{n_-}_{n_+}[[\p]]$.
\end{enumerate}}

The rest of the paper consists of the proof of this Theorem.

\sectionn{The case $n_-=2k$}

{}From Theorem~3 of~\cite{Cohom} it follows that the general solution of
equation~(\ref{1}) for the first-order deformation is given by
\begin{equation}
C_1(z|f,g)=\frac{1}{6}\kappa^2_1P(z|f,g)+Q_{\zeta_1}(z|f,g)+c_{31}R(z|f,g)
+d_1^{\rm ad}t_{D1}(z|f,g), \label{6.7}
\end{equation}
\[
\varepsilon(\kappa_1)=\varepsilon(c_{31})=\varepsilon(\zeta_{C1}(z))=0,
\]
where $Q(z|f,g)$ and $R(z|f,g)$ are given by (\ref{Q}) and~(\ref{R})
respectively and $P(z|f,g)$ is defined by the relation
\[
P(z|f,g)=f(z)\left(\frac{\overleftarrow{\partial}}{\partial z^A} \omega^{AB}
\frac{\partial}{\partial z^B}\right)^3g(z),\quad \bar{P}(|f,g)=0.
\]
 The
forms $P(z|f,g)$, $Q(z|f,g)$, and $R(z|f,g)$ satisfy the relations
\[
\bar{Q}_\zeta(|f,g)=0,\quad \bar{R}(|f,g)=0,
\]
\[
\sigma(f,h)[P(z|f,Q_\zeta(|g,h))+Q_\zeta(z|f,P(|g,h))]+{\rm cycle}(f,g,h)=
\]
\begin{equation}
=\sigma(f,h)S_{\zeta_1}(z|f,g)\bar{h}+{\rm cycle}(f,g,h)=
-\sigma(f,h)d_2^{\rm ad}J_\zeta(z|f,g,h), \label{6.7aa}
\end{equation}
\[
S_\zeta(z|f,g)=\left( \frac{\overleftarrow{\partial}}{\partial
z^A}\omega^{AB} \frac{\partial}{\partial z^B}\right)^3\{g(z),\zeta_C(z)\}-
\]
\[
-\sigma(f,g) g(z)\left(\frac{\overleftarrow{\partial}}{\partial
z^A}\omega^{AB} \frac{\partial}{\partial z^B}\right)^3\{f(z),\zeta_C(z)\}-
\{f(z)\left(\frac{\overleftarrow{\partial}}{\partial z^A}\omega^{AB}
\frac{\partial}{\partial z^B}\right)^3g(z),\zeta_C(z)\},
\]
\begin{equation}
J_\zeta(z|f,g)=f(z)\left( \frac{\overleftarrow{\partial}}{\partial
z^A}\omega^{AB} \frac{\partial}{\partial z^B}\right)^3\zeta_C(z)\bar{g}-
\sigma(f,g)g(z)\left(\frac{\overleftarrow{\partial}}{\partial z^A}\omega^{AB}
\frac{\partial}{\partial z^B}\right)^3\zeta_C(z)\bar{f}, \label{6.7ab}
\end{equation}
\[
 \sigma(f,h)[P(z|f,R(|g,h))+R(z|f,P(|g,h))]+{\rm
cycle}(f,g,h)=
\]
\[
-2[P(z|f,g)\bar{h}+{\rm cycle}(f,g,h)] \equiv -2T_{PR}(z|f,g,h),
\]
\begin{equation}
\sigma(f,h)Q_\zeta(z|f,Q_\zeta(|g,h))+{\rm cycle}(f,g,h)=0, \label{6.7a}
\end{equation}
\begin{equation}
\sigma(f,h)[Q_\zeta(z|f,R(|g,h))+R(z|f,Q_\zeta(|g,h))]+{\rm cycle}(f,g,h)=0,
\label{6.7b}
\end{equation}
\begin{equation}
\sigma(f,h)R(z|f,R(|g,h))+{\rm
cycle}(f,g,h)=0. \label{6.7c}
\end{equation}
In deriving these formulas, we
have used the identity
\[
\hat{l}[f(z)\left(\frac{\overleftarrow{\partial}}{\partial z^A}\omega^{AB}
\frac{\partial}{\partial z^B}\right)^pg(z)]=[\hat{l}f(z)]
\left(\frac{\overleftarrow{\partial}}{\partial z^A}\omega^{AB}
\frac{\partial}{\partial z^B}\right)^pg(z)+
f(z)\left(\frac{\overleftarrow{\partial}}{\partial z^A}\omega^{AB}
\frac{\partial}{\partial z^B}\right)^p\hat{l}g(z)+
\]
\[
+(p-1)f(z)\left(\frac{\overleftarrow{\partial}}{\partial z^A}\omega^{AB}
\frac{\partial}{\partial z^B}\right)^pg(z),\quad
\hat{l}=1-\frac{1}{2}z^A\frac{\partial}{\partial z^A}.
\]
Performing the
equivalence transformation $C(z|f,g)\rightarrow C_{T}(z|f,g)$ with
$T(z|f)=f(z)-\hbar^2t_{D1}(z|f)+O(\hbar^4)$, we rewrite $C(z|f,g)$ in the
form
\[
C(z|f,g)={\cal M}_{\kappa_1}(z|f,g)+\hbar^2Q_{\zeta_1}(z|f,g)+
c_{31}\hbar^2R(z|f,g)+\hbar^4C_2(z|f,g)+O(\hbar^6),
\]
where ${\cal
M}_\kappa(z|f,g)$ is the Moyal superbracket~(\ref{2}).

By the Jacobi identity~(\ref{6.1a}) for $C(z|f,g,h)$, we obtain
\begin{equation}
d_2^{\rm
ad}C_2(z|f,g,h)=\frac{c_{31}\kappa_1^2}{3}\sigma(f,h)T_{PR}(z|f,g,h)
+\frac{\kappa^2_1}{6}d_2^{\rm ad}J_{\zeta_1}(z|f,g,h). \label{6.8}
\end{equation}

Let
\[
\left[z\bigcup{\rm supp}(f)\bigcup{\rm supp}(g)\bigcup{\rm
supp}(h)\right] \bigcap{\rm supp}(s)=\emptyset;
\]
{}From equation (\ref{6.8})
it follows that
\begin{equation}
d_1^{\rm  ad}\hat{C}_2(z|f,g,h)=
\frac{c_{31}\kappa_1^2}{3}\sigma(f,h)P(z|f,g)\bar{h}+
\frac{\kappa^2_1}{6}d_1^{\rm ad}j_{\zeta_1}(z|f,g)\bar{h}, \label{6.9}
\end{equation}
where in the left-hand side of equation~(\ref{6.9}) the
differential $d_1^{\rm ad}$ acts only on the first argument of the functional
$\hat{C}_2(f,h)$ and
\[
j_{\zeta_1}(z|f)=f(z)\left(
\frac{\overleftarrow{\partial}}{\partial z^A}\omega^{AB}
\frac{\partial}{\partial z^B}\right)^3\zeta_C(z).
\]
As shown in Section~4.3
of~\cite{Cohom}, equation~(\ref{6.9}) has a solution only if
\[
c_{31}\kappa_1=0.
\]

i) $\kappa_1=0$, $c_{31}\neq0$.

In this case, equation~(\ref{6.8}) takes the form
\[
d_2^{\rm
ad}C_2(z|f,g,h)=0.
\]
The general solution of this equation is
\[
C_2(z|f,g)=\frac{1}{2}\kappa_2P(z|f,g)+Q_{\zeta_2}(z|f,g)+c_{32}R(z|f,g)
+d_1^{\rm ad}t_{D2}(z|f,g).
\]
Performing the equivalence transformation
$C(z|f,g)\rightarrow C_{T}(z|f,g)$ with
$T(z|f)=f(z)-\hbar^4t_{D2}(z|f,)+O(\hbar^6)$, we rewrite $C(z|f,g)$ in the
form
\[
C(z|f,g)=\{f(z),g(z)\}+\frac{1}{2}\hbar^4\kappa_2P(z|f,g)+
Q_{\zeta_{[2]}}(z|f,g)+c_{3[2]}R(z|f,g)+\hbar^6C_3(z|f,g)+O(\hbar^8),
\]
\[
\zeta_{[n]}(z)=\sum_{k=1}^n\hbar^{2k}\zeta_{Ck}(z),\quad
c_{3[n]}=\sum_{k=1}^n\hbar^{2k}c_{3k}.
\]
The Jacobi identity~(\ref{6.1a})
for $C(z|f,g,h)$ yields \begin{equation} d_2^{\rm
ad}C_3(z|f,g,h)=c_{31}\kappa_2\sigma(f,h)T_{PR}(z|f,g,h)+
\frac{\kappa_2}{2}d_2^{\rm ad}J_{\zeta_1}(z|f,g,h). \label{6.10}
\end{equation} We now conclude that equation~(\ref{6.10}) has a solution only
for $c_{31}\kappa_2=0$. This implies that $\kappa_2=0$.  Proceeding in the
same way, we find that up to an equivalence transformation, the general form
of the functional $C(z|f,g)$ satisfying the Jacobi identity~(\ref{6.1a}) (the
general form of the $*$-commutator) is given by \begin{equation}
C(z|f,g)=\{f(z),g(z)\}+Q_{\zeta_{[\infty]}}(z|f,g)+ c_{3[\infty]}R(z|f,g),
\label{6.11} \end{equation} A direct check using the relations~(\ref{6.7a}) -
(\ref{6.7c}) shows that the form~(\ref{6.11}) satisfies the Jacobi
identity~(\ref{6.1a}).

ii) $\kappa_1\neq0$, $c_{31}=0$.

In this case, equation~(\ref{6.8}) takes the form
\[
d_2^{\rm
ad}C_2(z|f,g,h)= \frac{\kappa^2_1}{6}d_2^{\rm ad}J_{\zeta_1}(z|f,g,h).
\]
The general solution of this equation is given by
\[
C_2(z|f,g)=\frac{\kappa^2_1}{6}J_{\zeta_1}(z|f,g)+
\frac{\kappa_1\kappa_2}{3}P(z|f,g)+Q_{\zeta_2}(z|f,g)+c_{32}R(z|f,g)
+d_1^{\rm ad}t_{D2}(z|f,g).
\]
Performing the equivalence transformation
$C(z|f,g)\rightarrow C_{T}(z|f,g)$ with
$T(z|f)=f(z)-\hbar^4t_{D2}(z|f,)+O(\hbar^6)$, we represent $C(z|f,g)$ in the
form \begin{equation} C(z|f,g)={\cal N}_{\kappa_{[2]},\zeta_{[2]}}(z|f,g)+
c_{32}\hbar^4R(z|f,g)+\hbar^6C_3(z|f,g)+O(\hbar^8), \label{6.12}
\end{equation} where\footnote{In deriving this relation, we have used the
equality
\[
\zeta(z)\left(\frac{\overleftarrow{\partial}}{\partial
z^A}\omega^{AB} \frac{\partial}{\partial z^B}\right)^p\zeta(z)=0
\]
which
holds for $\varepsilon(\zeta)=0$, $p=2k+1$.}
\[
{\cal
N}_{\kappa,\zeta}(z|f,g)={\cal M}_\kappa(z|f,g)+ {\cal
M}_\kappa(z|f,\zeta_C)\bar{g}- \sigma(f,g){\cal
M}_\kappa(z|g,\zeta_C)\bar{f})= {\cal
M}_\kappa(z|f+\zeta_C\bar{f},g+\zeta_C\bar{g}),
\]
\[
\kappa_{[n]}=\sum_{k=1}^n\hbar^{2(k-1)}\kappa_k.
\]
The form ${\cal
N}_{\kappa,\zeta}(z|f,g)$ satisfies the Jacobi identity~(\ref{6.1a}).  The
proof of this fact becomes obvious if the relation
\[
\int
dzf(z)\left(\frac{\overleftarrow{\partial}}{\partial z^A}\omega^{AB}
\frac{\partial}{\partial z^B}\right)^p\zeta(z)=0,\quad p\ge1.
\]
is taken
into account. The Jacobi identity~(\ref{6.1a}) for $C(z|f,g,h)$ yields
\begin{equation} d_2^{\rm
ad}C_3(z|f,g,h)=\frac{c_{32}\kappa_1^2}{3}\sigma(f,h)T_{PR}(z|f,g,h).
\label{6.13} \end{equation} This equation has a solution only for
$c_{32}\kappa_1=0$, whence it follows that $c_{32}=0$.

Proceeding in the same way, we find that up to an equivalence transformation,
the general form of the functional $C(z|f,g)$ satisfying the Jacobi
identity~(\ref{6.1a}) (the general form of the $*$-commutator) is given by
\[
C(z|f,g)={\cal N}_{\kappa_{[\infty]},\zeta_{[\infty]}}(z|f,g),\quad
\zeta_{C\infty}(z)\in{\raise2pt\hbox{$\mathbf
E^{n_-}_{n_+}$}}\big/{\raise-2pt\hbox{$\mathbf D^{n_-}_{n_+}$}},
\;\varepsilon(\zeta_{C\infty})=0.
\]

\sectionn{The case $n_-=2k+1$}

In this case, the general solution of equation~(\ref{1}) for the first-order
deformation is given by
\[
C_1(z|f,g)=\frac{1}{6}\kappa^2_1P(z|f,g)+Q_{\zeta_1}(z|f,g) +d_1^{\rm
ad}t_{D1}(z|f,g),
\]
\[
\varepsilon(\kappa_1)=0,\quad
\varepsilon(\zeta_{C1}(z))=1.
\]
Note that in contrast to~(\ref{6.7}), the
term $R(z|f,g)$ does not enter in the first-order deformation because for odd
$n_-$, it does not possess the required parity.

Consider the form
\[
{\cal N}_{\kappa,\zeta,c_2}(z|f,g)= {\cal
M}_\kappa(z|f+\zeta_C\bar{f},g+\zeta_C\bar{g})+c_2\bar{f}\bar{g}=
\]
\[
={\cal M}_\kappa(z|f,g)+ {\cal M}_\kappa(z|f,\zeta_C)\bar{g}-
\sigma(f,g){\cal M}_\kappa(z|g,\zeta_C)\bar{f}+ {\cal
L}_{\kappa,\zeta,c_2}(z)\bar{f}\bar{g},
\]
\[
{\cal
L}_{\kappa,\zeta,c_2}(z)={\cal M}_\kappa(z|\zeta_C,\zeta_C)+c_2,\quad
\varepsilon(\zeta_C(z))=1.
\]
We assume that $\zeta_C(z)\in \mathbf
E^{n_-}_{n_+}$ and ${\cal L}_{\kappa,\zeta,c_2}(z)\in \mathbf D^{n_-}_{n_+}$.
The form ${\cal N}_{\kappa,\zeta,c_2}(z|f,g)$ has the following properties:

i) ${\cal N}_{\kappa,\zeta,c_2}(z|f,g)\in \mathbf D^{n_-}_{n_+}$ (obvious).

ii) $\bar{{\cal N}}_{\kappa,\zeta,c_2}(f,g)=\int dz {\cal
N}_{\kappa,\zeta,c_2}(z|f,g)=0$.

Indeed,
\[
\int dz{\cal N}_{\kappa,\zeta,c_2}(z|f,g)=[\int dz {\cal
L}_{\kappa,\zeta,c_2}(z)]\bar{f}\bar{g}.
\]
The integral of ${\cal
L}_{\kappa,\zeta,c_2}(z)\in \mathbf D^{n_-}_{n_+}$ is well-defined and is
equal to zero as an integral of an even function with respect to an odd
number of odd variables.

iii) The form ${\cal N}_{\kappa,\zeta,c_2}(z|f,g)$ satisfies the Jacobi
identity.

Indeed,
\[
{\cal N}_{\kappa,\zeta,c_2}(z|f,{\cal
N}_{\kappa,\zeta,c_2}(|g,h))= {\cal M}_\kappa(z|f+\zeta_C\bar{f},{\cal
M}_\kappa(|g+\zeta_C\bar{g}, h+\zeta\bar{h}))+{\cal
M}_\kappa(z|f+\zeta_C\bar{f},c_2\bar{g}\bar{h})+
\]
\[
+[{\cal
M}_\kappa(z|f+\zeta_C\bar{f},\zeta_C)+c_2\bar{f}] \bar{{\cal
N}}_{\kappa,\zeta,c_2}(g,h)= {\cal M}_\kappa(z|f+\zeta_C\bar{f},{\cal
M}_\kappa(|g+\zeta_C\bar{g}, h+\zeta\bar{h})),
\]
whence the Jacobi identity
follows because the form ${\cal M}_\kappa(z|f,g)$ satisfies the Jacobi
identity for any $f,g,h\in \mathbf E^{n_-}_{n_+}$.

Now consider the form
\[
{\cal N}_{\kappa_{[n]},\zeta_{[n]},c_{2[n]}}(z|f,g)=
{\cal M}_{\kappa_{[n]}}(z|f+\zeta_{C[n]}\bar{f},g+\zeta_{C[n]}\bar{g})+
c_{2[n]}\bar{f}\bar{g}=
\]
\[
={\cal M}_{\kappa_{[n]}}(z|f,g)+ {\cal
M}_{\kappa_{[n]}}(z|f,\zeta_{C[n]})\bar{g}- \sigma(f,g){\cal
M}_{\kappa_{[n]}}(z|g,\zeta_{C[n]})\bar{f}+ {\cal
L}_{\kappa_{[n]},\zeta_{[n]},c_{2[n]}}(z)\bar{f}\bar{g},\quad n\ge1,
\]
where
\[
\kappa_{[n]}=\!\sum_{k=1}^n\hbar^{2(k-1)}\kappa_k,\quad
\zeta_{C[n]}(z)=\!\sum_{k=1}^n\hbar^{2k)}\zeta_{Ck}(z),\;
\varepsilon(\zeta_{Ck}(z))=1,\quad
c_{2[n]}=\!\sum_{k=1}^n\hbar^{2k)}c_{2k},\; c_{21}=0.
\]
Suppose that
\[
\zeta_{C[n]}\in C^\infty,\quad {\cal
L}_{\kappa_{[n]},\zeta_{[n]},c_{2[n]}}\biggr|_{[n+1]}(z)\in \mathbf
D^{n_-}_{n_+},
\]
where
$A(\hbar^2)\biggr|_{[n]}=\sum\limits_{k=0}^nA(\hbar^2)\biggr|_k$, and
$A(\hbar^2)\biggr|_k$ denotes the term of order $\hbar^{2k}$ of the expansion
of the function $A(\hbar^2)$ in the Taylor series in $\hbar^2$.  Then it is
easy to show that the Jacobian $J{\cal
N}_{\kappa_{[n]},\zeta_{[n]},c_{2[n]}}(z|f,g,h)$ of the form ${\cal
N}_{\kappa_{[n]},\zeta_{[n]},c_{2[n]}}(z|f,g)$,
\[
J{\cal
N}_{\kappa_{[n]},\zeta_{[n]},c_{2[n]}}(z|f,g,h)\equiv\sigma(f,h) {\cal
N}_{\kappa_{[n]},\zeta_{[n]},c_{2[n]}}(z|f, {\cal
N}_{\kappa_{[n]},\zeta_{[n]},c_{2[n]}}(|g,h))+{\rm cycle}(f,g,h),
\]
has the
property
\[
J{\cal
N}_{\kappa_{[n]},\zeta_{[n]},c_{2[n]}}(z|f,g,h)\biggr|_{[n+2]}=0
\]
(in this
Jacobian, the terms of order $\hbar^{2k}$, $k\ge n+3$, may be ill-defined).

Performing the equivalence transformation $C(z|f,g)\rightarrow C_{T}(z|f,g)$
with $T(z|f)=f(z)-\hbar^2t_{D1}(z|f)+O(\hbar^4)$, we rewrite $C(z|f,g)$ in
the form
\[
C(z|f,g)={\cal N}_{\kappa_{[1]},\zeta_{[1]},c_{2[1]}}(z|f,g)+
\hbar^4C_2(z|f,g)- {\cal
L}_{\kappa_{[1]},\zeta_{[1]},c_{2[1]}}(z)\biggr|_2\bar{f}\bar{g}+ O(\hbar^6).
\]
The condition $C(z|f,g)\in \mathbf D^{n_-}_{n_+}$ implies that
$C_2(z|f,g)\in \mathbf D^{n_-}_{n_+}$.

The Jacobi identity~(\ref{6.1a}) for $C(z|f,g,h)$ yields
\[
d_2^{\rm
ad}C'_2(z|f,g,h)=0,\quad C'_2(z|f,g)=C_2(z|f,g)-\hbar^{-4} {\cal
L}_{\kappa_{[1]},\zeta_{[1]},c_{2[1]}}(z)\biggr|_2\bar{f}\bar{g},
\]
and
hence
\[
C_2(z|f,g)=\hbar^{-4} {\cal
L}_{\kappa_{[1]},\zeta_{[1]},c_{2[1]}}(z)\biggr|_2\bar{f}\bar{g}+
\frac{\kappa_1\kappa_2}{3}P(z|f,g)+c_{22}\bar{f}\bar{g}+
\]
\[
+\delta_{n_-,4+n_+}c_{42}V(z|f,g)+d_1^{\rm ad}t_2(z|f,g),
\]
where
\[
V(z|f,g)=\int duu^A\left(\frac{\partial f(u)}{\partial u^A}g(u)-
\sigma(f,g)\frac{\partial g(u)}{\partial u^A}f(u)\right).
\]

Let
\[
z\bigcup{\rm supp}(f)=z\bigcup{\rm supp}(g)= {\rm supp}(f)\bigcup{\rm
supp}(g)=\emptyset.
\]
Then we have
\[
\hat{C}_2(z|f,g)=[\hbar^{-4} {\cal
L}_{\kappa_{[1]},\zeta_{[1]},c_{2[1]}}(z)\biggr|_2+ c_{22}]\bar{f}\bar{g}.
\]
The condition $\hat{C}_2(z|f,g)\in \mathbf D^{n_-}_{n_+}$ yields
\begin{equation} [\hbar^{-4} {\cal
L}_{\kappa_{[1]},\zeta_{[1]},c_{2[1]}}(z)\biggr|_2+ c_{22}]\in \mathbf
D^{n_-}_{n_+}. \label{6.14} \end{equation} Thus, the deformation is possible
not for all functions $\zeta_{C1}(z)$, but only for those satisfying the
condition~(\ref{6.14}).

Further, from the condition $[\delta_{n_-,4+n_+}c_{42}V(z|f,g)+d_1^{\rm
ad}t_2(z|f,g)] \in \mathbf D^{n_-}_{n_+}$ we find (see the end of Section 4.3
of~\cite{Cohom}) that
\[
c_{42}=0,\quad
t_2(z|f)=\zeta_{C2}(z)\bar{f}+t_{D2}(z|f),
\]
\[
\zeta_{C2}(z)\in{\raise2pt\hbox{$\mathbf
E^{n_-}_{n_+}$}}\big/{\raise-2pt\hbox{$\mathbf D^{n_-}_{n_+}$}},\quad
\varepsilon(\zeta_{C2})=1,\quad t_{D2}(z|f)\in \mathbf D^{n_-}_{n_+}.
\]

Performing the equivalence transformation $C(z|f,g)\rightarrow C_{T}(z|f,g)$
with $T(z|f)=f(z)-\hbar^4t_{D2}(z|f)+O(\hbar^6)$, we represent $C(z|f,g)$ in
the form
\[
C(z|f,g)={\cal N}_{\kappa_{[2]},\zeta_{[2]},c_{2[2]}}(z|f,g)+
\hbar^6C_3(z|f,g)- {\cal
L}_{\kappa_{[2]},\zeta_{[2]},c_{2[2]}}(z)\biggr|_3\bar{f}\bar{g}+ O(\hbar^8).
\]
The condition $C(z|f,g)\in \mathbf D^{n_-}_{n_+}$ implies that
$C_3(z|f,g)\in D$.

The Jacobi identity~(\ref{6.1a}) for $C(z|f,g,h)$ yields
\[
d_2^{\rm
ad}C'_3(z|f,g,h)=0,\quad C'_3(z|f,g)=C_3(z|f,g)-\hbar^{-6} {\cal
L}_{\kappa_{[2]},\zeta_{[2]},c_{2[2]}}(z)\biggr|_3\bar{f}\bar{g},
\]
whence
it follows that
\[
C_3(z|f,g)=\hbar^{-6} {\cal
L}_{\kappa_{[2]},\zeta_{[2]},c_{2[2]}}(z)\biggr|_3\bar{f}\bar{g}+
(\frac{\kappa_2^2}{6}+\frac{\kappa_1\kappa_3}{3})P(z|f,g)+
c_{23}\bar{f}\bar{g}+
\]
\[
+\delta_{n_-,4+n_+}c_{43}V(z|f,g)+d_1^{\rm
ad}t_3(z|f,g),
\]

Let
\[
z\bigcup{\rm supp}(f)=z\bigcup{\rm supp}(g)= {\rm supp}(f)\bigcup{\rm
supp}(g)=\emptyset.
\]
Then we have
\[
\hat{C}_3(z|f,g)=[\hbar^{-6} {\cal
L}_{\kappa_{[2]},\zeta_{[2]},c_{2[2]}}(z)\biggr|_3+ c_{23}]\bar{f}\bar{g}.
\]
The condition $\hat{C}_3(z|f,g)\in \mathbf D^{n_-}_{n_+}$ gives
\[
[\hbar^{-6} {\cal
L}_{\kappa_{[2]},\zeta_{[2]},c_{2[2]}}(z)\biggr|_3+c_{22}]\in \mathbf
D^{n_-}_{n_+}.
\]

Proceeding in the same way, we finally find that up to an equivalence
transformation, the general form of the functional $C(z|f,g)$ satisfying the
Jacobi identity~(\ref{6.1a}) (the general form of the $*$-commutator) is
given by
\[
C(z|f,g)={\cal
N}_{\kappa_{[\infty]},\zeta_{[\infty]},c_{2[\infty]}}(z|f,g),
\]
\[
\zeta_{C\infty}(z)\in{\raise2pt\hbox{$\mathbf
E^{n_-}_{n_+}$}}\big/{\raise-2pt\hbox{$\mathbf D^{n_-}_{n_+}$}},
\;\varepsilon(\zeta_{C\infty})=1,\quad [{\cal
M}_{\kappa_{[\infty]}}(z|\zeta_{C[\infty]},\zeta_{C[\infty]})+
c_{2[\infty]}]\in \mathbf D^{n_-}_{n_+}.
\]

\sectionn{Acknowledgements} This work was supported by the RFBR (grants
               No.~02-02-16944 (I.T.), No.~02-02-17067 (S.K.), and
               No.~02-02-16946 (A.S.)), by INTAS (grant No.~03-51-6346 (A.S.)
               and No.~00-00-262 (I.T.)) and by the grant LSS-1578.2003.2.

\end{document}